\documentclass[12pt]{article}
\usepackage[pdftex]{graphicx}
\usepackage{hyperref}
\usepackage{amsmath}
\usepackage{amssymb}
\renewcommand{\title}[1]{%
    \bigskip%
    \begin{center}%
    \Large\bf #1%
    \end{center}%
    \vskip .2in}
\renewcommand{\author}[1]{%
    {\begin{center}
    #1
    \end{center}}}

\begin{document}

\title{\bf{Dual Description of Gauge Theories from an Iterative Noether Approach}}

\centerline{{Rabin Banerjee}\footnote {rabin@bose.res.in}}

\centerline{S. N. Bose National Centre 
for Basic Sciences
, JD Block, Sector III, Salt Lake City,} 
\centerline{Kolkata -700 106, India }

\vskip 1cm
\begin{abstract}
An iterative Noether scheme, advocated by Deser, is used to introduce gauge invariant couplings to nonrelativistic matter with  global symmetries related to usual charge conservation and dipole conservation recently discussed in fractonic theories. No reference to any gauge principle, fractonic or otherwise, is required. A dual description is found where the theory is defined either in terms of the usual vector gauge field or, alternatively, in terms of higher derivatives of a symmetric tensor field given in fractonic theories. A connection between these two descriptions is obtained by providing an explicit map between the vector and tensor fields. This method yields a novel `minimal' prescription involving  tensor fields which is identified with the usual minimal prescription that involves  vector fields, by using the mapping.   It also spells out the structure of the pure gauge field action  in  both formulations. Extension of the abelian $U(1)$ invariance to the nonabelian $SU(N)$ invariance is done for the standard formulation.
\end{abstract}

\section{Introduction}
\smallskip

The recent study of a new class of nonrelativistic symmetric higher rank tensor gauge theories {\cite{P1, P2, DRH, S}} has led to the intriguing possibility of the existence of novel topological excitations, called fractons. Their analysis has found applications in various condensed matter contexts {\cite{C, NH, PCY, QRH, SS}}  and even in gravitational physics {\cite{SPP, Umang, RB}}. They have the unusual property of restricted motion which may vary from complete immobility to mobility in subdimensional spaces.  This is a consequence of nonstandard conservation laws since the matter sector admits general global symmetries which, apart from the usual charge symmetry, also has dipole symmetries. This makes even the  pure matter action (`free' fractonic theory) non-gaussian, leading to nonlinear equations of motion. The matter action is constructed up to leading orders in the derivatives by using the specific structure of the charge and dipole moment symmetries. Interactions are introduced by exploiting the fractonic gauge principle {\cite{Pretko}} which demands that the original invariance of the pure matter sector under global symmetries is retained under the local transformations. This requires the introduction of a new symmetric higher rank tensor (fractonic) field with a specific transformation property. 

Simultaneously,  in a parallel but related development, it was shown {\cite{S}} that interactions may also be introduced using the usual vector field where the gauge principle now ensures that all ordinary derivatives are replaced by their covariant expressions. Thus there are two competing approaches, based on usual vector fields (standard formulation) or on symmetric tensor fields (nonstandard formulation).

Long back Deser {\cite{Deser}} developed an approach where interactions are inducted by an iterative Noether prescription without any need for a gauge principle. This method, which emphasises the role of interactions, proceeds by adding to the original theory a source that is obtained from the free part itself, followed by the addition of a further source due to this one and so on till the iteration stops due to the lack of any new source. It was originally applied to obtain the $SU(2)$ Yang-Mills theory (a nonlinear theory) from the linear Maxwell theory where self interactions were suitably inducted. Several examples of this approach can be found including an extension to matter fields whose interactions with gauge fields naturally lead to the minimal coupling {\cite{DHB, AS, DH, K, K1, BMM}}. All these applications consider relativistic theories where the basic starting lagrangian is quadratic with linear equations of motion.

In this paper we use this formalism in a completely different context, namely, nonrelativistic theories where the pure matter sector is nonlinear. It is therefore a new application of the method which has to be generalised to incorporate the fact that the starting lagrangian is not quadratic and also includes higher derivatives. The purpose of the paper, however, is not merely to provide a new (and somewhat nontrivial) application. More importantly, it yields fresh insights in the construction of interacting gauge theories beginning from a matter sector that admits generalised global symmetry yielding conservation of both charge and dipole moment. 

Starting  from a pure matter sector that has both these symmetries, the implementation of the formalism leads to a dual description of interacting gauge theories. In one scenario the vector current is gauged with the usual vector field and the interacting theory turns out to be the one where all ordinary derivatives are replaced by their covariant derivatives. Thus the standard minimal coupling prescription obtained by the gauge principle is reproduced. However there is another possibility. The consequence of two global symmetries necessarily leads to a construction where the vector current is expressed as the gradient of a second rank symmetric tensor current. This comes out naturally in the Noether prescription. Instead of considering the vector current as the source, the tensor current is taken as the source in the iterative scheme. This leads to the fractonic gauge theory recently considered in the literature {\cite{P1, P2, DRH, S, NH, RB}}. As already mentioned the two currents are related by a spatial derivative. This is used to find an exact map between the  vector field and the  tensor gauge field, thereby relating the two descriptions.  We further extend the analysis to include more global symmetries. The dual picture still holds, except that the relation connecting the gauge fields becomes different.

The iterative Noether scheme is introduced in section 2 where, starting from the free Schroedinger theory, the minimally coupled interacting theory is obtained without using any gauge principle.   Dual description of gauge theories, the central point of the paper, is treated in section 3. There is an overview of conservation laws related to  charge and dipole moment symmetries. Next, the matter sector is given and interactions are found by the method of iteration. This gives a dual description expressed either in terms of a standard vector field or, alternatively, in terms of a symmetric second rank fractonic tensor field. A connection between the two formulations is found. The interactions involving the tensor field can also be obtained by a novel `minimal prescription'. Its connection with the conventional minimal prescription is elucidated.  The explicit structures for the pure gauge field lagrangians are given. Finally, an extension to $SU(N)$ invariance is done. Section 4 discusses similar findings if other conservation laws besides charge and dipole moment  exist. Conclusions are given in section 5.

\section{ Schroedinger Theory from Noether Approach}

\smallskip
Here we show how interactions in the usual Schroedinger theory are introduced by adopting an iterative Noether prescription {\cite{Deser}} instead of resorting to any gauge principle. Consider the free Schroedinger theory for a complex scalar field,
\begin{equation}
{\cal L}_0= \frac{i}{2}\phi^* ({\stackrel{\leftrightarrow}{\partial}}_0) \phi -\frac{1}{2m}\partial_i\phi^* \partial_i\phi
\label{1}
\end{equation}
It has a global $U(1)$ symmetry,
\begin{equation}
    \phi\rightarrow e^{i\alpha}\phi\,\,;\,\, \phi^*\rightarrow e^{-i\alpha}\phi^*
    \label{2}
\end{equation}
which leads to the conservation law,
\begin{equation}
    \partial_0 j_0 - \partial_i j_i =0
    \label{3}
\end{equation}
from Noether's first theorem. This theorem yields the explicit structures for the currents,
\begin{equation}
    \alpha j^0=\frac{\partial{\cal L}_0}{\partial(\partial_0\phi)}\delta\phi +  c.c.\,\,;\,\, \alpha j^i=\frac{\partial{\cal L}_0}{\partial(\partial_i\phi)}\delta\phi + c.c.
    \label{4}
\end{equation}
where $c.c.$ stands for complex conjugate. Using   $\delta\phi=i\alpha \phi$ we obtain,
\begin{equation}
  j^0=-j_0= -\phi^*\phi \,\,\,;\,\,\, j^i=j_i= \frac{i}{2m}\Big( (\partial_i\phi)\phi^* - (\partial_i\phi^*)\phi\Big) 
  \label{5}
\end{equation}
    \label{6}

The standard way to introduce interactions is to make the transformation parameter $\alpha$ local and adopt the gauge principle which ensures invariance under the local symmetry by replacing the usual derivatives by covariant derivatives. Here we follow a different path.

We obtain the final lagrangian by an iterative process. The analysis is done separately for the time and space components. For the time component, the first correction to the lagrangian is obtained by adding the term,
\begin{equation}
    {\cal L}_1 = j^0 A_0 = -\phi^*\phi A_0
    \label{7}
\end{equation}
Since there are no derivatives, the Noether prescription (\ref{4}) does not yield anything and the iteration concludes. Thus the final lagrangian, for the time part, is given by,
\begin{equation}
    {\cal L}_{time} = {\cal L}_0 + {\cal L}_1 = \frac{i}{2}\phi^* ({\stackrel{\leftrightarrow}{\partial}}_0) \phi -\phi^*\phi A_0
    \label{8}
\end{equation}
which may also be expressed in the standard minimal coupling form,
\begin{equation}
    {\cal L}_{time} = \frac{i}{2}\Big(\phi^* D_0\phi - \phi(D_0\phi)^*\Big) ; \hspace{1cm}  D_0\phi = (\partial_0 + i A_0)\phi
    \label{9}
\end{equation}

We next repeat the analysis for the space part. The first correction is obvious,
\begin{equation}
    {\cal L}_1= j^i A_i = \frac{i}{2m}\Big( (\partial_i\phi)\phi^* - (\partial_i\phi^*)\phi\Big) A_i
    \label{10}
\end{equation}
Since derivatives occur, there is a next correction to be evaluated by the Noether prescription, replacing ${\cal L}_0$ by ${\cal L}_1$
in (\ref{4}). We find,
\begin{equation}
    \alpha j_{1}^i=\frac{\partial{\cal L}_1}{\partial(\partial_i\phi)}\delta\phi + c.c.
    \label{11}
\end{equation}
which yields,
\begin{equation}
    j_1^i= -\frac{1}{m}\phi^*\phi A_i
    \label{12}
\end{equation}
The contribution to the lagrangian is given by,
\begin{equation}
    {\cal L}_2= \frac{1}{2}j_1^i A_i = -\frac{1}{2m} \phi^*\phi A_i^2
\end{equation}
\label{13}
The half factor is necessary to reproduce the expression for the current from a variation of the action,
\begin{equation}
  j_1^i(x)=\frac{\delta}{\delta A_i(x)}\int\, {\cal L}_2
  \label{14}
\end{equation}
Since there are no derivatives, the iteration stops and the final lagrangian for the spatial part is given by adding the contributions,
\begin{equation}
    {\cal L}_{space}= {\cal L}_0 + {\cal L}_1 + {\cal L}_2 = -\frac{1}{2m}\partial_i\phi^* \partial_i\phi + \frac{i}{2m}\Big( (\partial_i\phi)\phi^* - (\partial_i\phi^*)\phi\Big) A_i  -\frac{1}{2m} \phi^*\phi A_i^2
    \label{15}
\end{equation}
which may be rearranged to display the minimal coupling form,
\begin{equation}
    {\cal L}_{space}= -\frac{1}{2m} (D_i\phi)^* (D_i\phi) \hspace{1cm}  D_i\phi= (\partial_i + i A_i)\phi
    \label{16}
\end{equation}
    \label{17}
    \label{18}

The complete interacting lagrangian, which is a sum of (\ref{9}) and (\ref{16}), is given by,
\begin{equation}
    {\cal L}_{I}={\cal L}_{time} + {\cal L}_{space} = \frac{i}{2}\Big(\phi^* D_0\phi - \phi(D_0\phi)^*\Big)  -\frac{1}{2m} (D_i\phi)^* (D_i\phi)
    \label{19}
\end{equation}
Expectedly, the above  lagrangian  is invariant under the local gauge transformations,
\begin{equation}
    \phi\rightarrow e^{i\alpha(x)\phi}\,\, ; \,\, A_\mu\rightarrow A_\mu - \partial_\mu \alpha\label{combo}
\end{equation}
The dynamics for the gauge field is given by the usual Maxwell type term,{\footnote{Contrary to the relativistic case, the  coefficients $C_1, C_2$, multiplying the electric and magnetic parts can be different.}}
\begin{equation}
    {\cal L}_{gauge}= C_1 F_{0i}^2 - C_2  F_{ij}^2\hspace{1cm} F_{0i}=\partial_0 A_i - \partial_i A_0 \hspace{0.5cm} F_{ij}= \partial_i A_j - \partial_j A_i
    \label{20}
\end{equation}
The extension to the nonabelian case may be done along similar lines, This analysis is deferred to section 3.2.3 when the higher derivative example is treated.




  \label{21}
    \label{22}

    \label{24}
    \label{25}
    \label{27}
    \label{28}

    \label{29}
    \label{30}
    \label{31}
    \label{32}
    \label{33}
    \label{34}
    \label{35}
    \label{36}
    \label{37}
    \label{38}
    \label{39}

  \label{41}
   \label{42}
   \label{43}
    \label{44}

 \label{45}
    \label{46}
    \label{47}


\section{Dual Description of  Gauge Theories}

Recently a new class of higher derivative gauge theories which exhibit, apart from  standard charge conservation, other  conservation laws related to the dipole symmetry have come to the fore. Their study has led to the discovery and understanding of a new type of topological matter called fractons.

After a general overview that discusses the various global symmetries  in fractonic theories, we show that the Noether prescription correctly reproduces the various conservation laws associated with them in a concrete example involving complex scalars. The iterative Noether approach is then applied to obtain the explicit couplings of these scalars to gauge fields. The higher derivative nature of the theory allows for either minimal or nonminimal couplings, leading to a dual description. 

\subsection{Overview of Symmetries and Conservation Laws}

Let us consider some matter theory that has, as sources, $J_0$ and a symmetric tensor $J_{ij}$ satisfying {\cite{S, RB}},
\begin{equation}
    \partial_o J_0 - \partial_i\partial_j J_{ij}=0
    \label{48}
\end{equation}
This can be achieved, for instance, by coupling these sources to $A^0$ and $A_{ij}$, respectively. Here $A_{ij}$ is a symmetric tensor gauge field that can, among others, characterise a fractonic gauge field that transforms as,
\begin{equation}
    \delta A_{ij}= \partial_i\partial_j \alpha\label{49}
\end{equation}
while $A_0$ satisfies the standard rule (\ref{combo}).
Gauge invariance under these transformations ensures the conservation law (\ref{48}).

We now analyse the global symmetries. There is the obvious ordinary global symmetry with currents,
\begin{equation}
    j_0=J_0 \hspace{1.5cm} j_i= \partial_j J_{ij}
    \label{51}
\end{equation}
satisfying,
\begin{equation}
    \partial_0 j_0 - \partial_i j_i=0
    \label{52}
\end{equation}
with the usual conserved charge,
\begin{equation}
    Q=\int_{space} j_0 = \int_{space} J_0
    \label{53}
\end{equation}
Further, it has a vector global symmetry with currents,
\begin{equation}
    j_0^i=x^i J_0 \hspace{1cm} j_{ij}= x^j \partial_k J^{ki} - J_{ij}
    \label{54}
\end{equation}
having the conservation law,
\begin{equation}
   \partial_0 j_0^i - \partial_j j_{ji} =0
   \label{55}
\end{equation}
The corresponding conserved charge is given by,
\begin{equation}
    Q^i = \int_{space} j_0^i = \int_{space} x^i J_0
    \label{56}
\end{equation}
\smallskip
The global symmetry (\ref{55}), (\ref{56}) is interpreted as a dipole symmetry
 just as the usual one (\ref{53}) is considered as a charge symmetry. It implies that a single charge cannot move although restricted mobility is allowed for a system of charges. These features are characteristic of  fractons.
 
 \subsection{Matter Sector and Interactions}
 
 An explicit construction of the matter sector in terms of complex scalars, satisfying the conservation laws (\ref{52}) and (\ref{55}), has been given in the literature. We briefly review this construction and then show how gauge couplings are introduced using the iterative Noether prescription. The couplings can be interpreted either in terms of a standard gauge field or, alternatively, in terms of a higher rank tensor field that occurs in fractonic theories.
 
 Since the theory obeys both charge and dipole conservation, the appropriate symmetry transformation on the complex scalar takes the form,
\begin{equation}
    \phi(x)\rightarrow e^{i\alpha(x)}\phi(x) \hspace{1cm} \alpha(x)=\alpha_0 + \alpha_i x^i
    \label{57}
\end{equation}
where $\alpha_0, \alpha_i$ are global (constant) parameters. The matter lagrangian invariant under the above transformations modifies the spatial derivative terms of (\ref{1}) and is given by \cite{Pretko, S},
\begin{equation}
    {\cal L}_0= \frac{i}{2}\phi^* ({\stackrel{\leftrightarrow}{\partial}}_0) \phi - s\,\partial_i(\phi^*\phi)\partial_i(\phi^*\phi) +t\, (i\phi^{*2}\partial_i\phi\partial^i\phi - i\phi^2\partial_i\phi^*\partial^i\phi^*) +u\, |{\phi\partial_i\partial_j\phi- \partial_i\phi\partial_j\phi}|^2 -\lambda |\phi|^4
    \label{58}
\end{equation}
where the leading order terms containing fourth powers of the field as well as the derivatives have only been retained.

The conserved currents are found by the Noether procedure. Noting that,
\begin{equation}
    \delta\phi=i\alpha \phi \hspace{1cm} \delta\phi^*= -i\alpha \phi^* \label{59}
\end{equation}
the time component of the current is given by,
\begin{equation}
    \alpha j^0=\frac{\partial{\cal L}_0}{\partial(\partial_0\phi)}\delta\phi + \frac{\partial{\cal L}_0}{\partial(\partial_0\phi^*)}\delta\phi^* =\alpha \phi^*\phi
    \label{60}
\end{equation}
so that,
\begin{equation}
    j_0=\phi^*\phi
    \label{61}
\end{equation}
This has the same form as conventional theories because there is no time dependence in the transformation parameter $\alpha$. 

For the computation of the space component, we consider the various terms one by one using the same definition as (\ref{4}). There is no contribution from the term involving $s$. For the $t$ term the result is,
\begin{equation}
    \alpha_0 j^i = \frac{\partial{\cal L}_0}{\partial(\partial_i\phi)}\delta\phi + \frac{\partial{\cal L}_0}{\partial(\partial_i\phi^*)}\delta\phi^* = -\alpha_0\, \partial_i|\phi|^4
    \label{62}
\end{equation}
so that,
\begin{equation}
j_i=- \, \partial_i|\phi|^4
\label{63}
\end{equation}
This can be put in the form $j_i= \partial_j J_{ij}$ (see \ref{51}), where,
\begin{equation}
   J_{ij}= - \delta_{ij} |\phi|^4
   \label{64}
\end{equation}
This relation was earlier given in \cite{S} by using a polar decomposition of the fractonic field that simplified the algebra.
Thus the general analysis in section 3.1 goes through and both conservation laws (\ref{52}) and (\ref{55}) hold. Observe that we have taken only the constant parameter $\alpha_0$ in (\ref{62}) to keep parity with the standard Noether definition.

The analysis for the $u$ term in (\ref{58}) is more involved since it involves double derivatives on the fields. In this case the derivation of the Noether current is based on the Euler-Lagrange equation valid up to second order derivatives which is given by, 
\begin{equation}
    \frac{{\partial}{\cal L}}{\partial\phi} -\partial_\mu\frac{{\partial}{\cal L}}{\partial(\partial_\mu\phi)} +\partial_\mu\partial_\nu\frac{{\partial}{\cal L}}{\partial(\partial_\mu\partial_\nu\phi)} = 0
    \label{el2}
\end{equation}
This leads to the Noether current, which is a generalization of (\ref{4}),
\begin{eqnarray}
    \alpha_0  j^i &=& \frac{\partial{\cal L}_0}{\partial(\partial_i\phi)}\delta\phi + \frac{\partial{\cal L}_0}{\partial(\partial_i\partial_j\phi)}\partial_j(\delta\phi) - \Big(\partial_j\frac{\partial{\cal L}_0}{\partial(\partial_i\partial_j\phi)}\Big)(\delta\phi)\cr &+& \frac{\partial{\cal L}_0}{\partial(\partial_i\phi^*)}\delta\phi^*  + \frac{\partial{\cal L}_0}{\partial(\partial_i\partial_j\phi^*)}\partial_j(\delta\phi^*) - \Big(\partial_j\frac{\partial{\cal L}_0}{\partial(\partial_i\partial_j\phi^*)}\Big)(\delta\phi^*)
    \label{65}
\end{eqnarray}

Substituting the $u$-term from (\ref{58}), keeping terms proportional to $\alpha_0$,  yields the following result,
\begin{eqnarray}
  \alpha_0  j^i = -i\alpha_0  \,  \partial_j\Big(\phi^2(\phi^*\partial_i\partial_j\phi^* -\partial_i\phi^*\partial_j\phi^*) -\phi^{*2}(\phi\partial_i\partial_j\phi -\partial_i\phi\partial_j\phi) \Big)
\end{eqnarray}
Equating terms proportional to $\alpha_0$, we find,
\begin{equation}
j^i= \, \partial_j J_{ij}\label{67}
\end{equation}
where,
\begin{equation}
    J_{ij}= -i\Big(\phi^2(\phi^*\partial_i\partial_j\phi^* -\partial_i\phi^*\partial_j\phi^*) -\phi^{*2}(\phi\partial_i\partial_j\phi -\partial_i\phi\partial_j\phi)\Big)\label{68}
\end{equation}
which reproduces the structure (\ref{51}) from the general discussion. Incidentally, this relation appeared earlier in \cite{S} by using equations of motion in polar form.

\subsubsection{Gauge Couplings}

In this section we discuss the introduction of gauge couplings to the matter fields by adopting the iterative Noether prescription \cite{Deser}. This method, successful in conventional theories, also holds here, leading to new insights. Especially, it provides a minimal coupling prescription for tensor gauge fields encountered here. Also, this analysis is used to give a map that connects the tensor field with the usual vector field.

We begin from the lagrangian (\ref{58}) and first consider the temporal part. This has the same form as standard theories, eventually leading to the minimally coupled form (\ref{9}).

The remaining terms involving spatial derivatives are now considered. For the $s$ term, as already stated there is no contribution to the Noether current. Thus there is no coupling associated with this term. To see that the same conclusion also follows by adopting the minimal prescription, the ordinary derivatives are replaced by covariant derivatives. The result is,
\begin{equation}
    \Big((\partial_i -iA_i)\phi^*\phi + \phi^*(\partial_i+iA_i)\phi\Big) \Big((\partial_i -iA_i)\phi^*\phi + \phi^*(\partial_i+iA_i)\phi\Big) \label{75}
\end{equation}
The $A_i$ terms in each of the big brackets cancel out leaving just the original terms. Thus there is no effect of the minimal coupling. This came out naturally in the Noether scheme.

We next consider the $t$ term in (\ref{58}). The current for this term was already computed in (\ref{63}). Thus the new addition in the lagrangian is given by,
\begin{equation}
   {\cal L}_1= j_i A_i =- \, \partial_i|\phi|^4 A_i
   \label{76}
\end{equation}
The iterative Noether process continues since the lagrangian involves derivatives. The corresponding current is obtained, as usual, by replacing ${\cal L}_0$ in (\ref{62}) by ${\cal L}_1$,
\begin{equation}
    j^i = \frac{\partial{\cal L}_1}{\partial(\partial_i\phi)}\delta\phi + \frac{\partial{\cal L}_1}{\partial(\partial_i\phi^*)}\delta\phi^* = 0
    \label{77}
\end{equation}
There is no contribution to the current and hence there is no correction to the lagrangian. Also, the iteration terminates and the final expression for the $t$ term is given by,
\begin{equation}
    {\cal L} = {\cal L}_0 + {\cal L}_1 = (i\phi^{*2}\partial_i\phi\partial^i\phi - i\phi^2\partial_i\phi^*\partial^i\phi^*) - \partial_i|\phi|^4 A_i\label{78}
\end{equation}
In this form  invariance is realised by the fact that the change in the pure matter sector under the local transformations ( where $\alpha(x)$ is arbitrary and not confined to the special form given in (\ref{57}) is precisely cancelled by the change in the coupling term. 

Although not obvious, the above lagrangian can be expressed in the standard minimally coupled form,
\begin{equation}
    {\cal L} =  i\phi^{*2}(\partial_i + iA_i)\phi\,(\partial^i +iA^i)\phi - i\phi^2(\partial_i - iA_i)\phi^*\,(\partial^i-iA^i)\phi^* 
    \label{79}
\end{equation}
Now gauge invariance under the local transformations 
becomes manifest,  exactly as happens for the usual case.

It is possible to rephrase the entire discussion in terms of nonminimal couplings and nonstandard gauge fields, recently introduced in the literature as fractonic gauge fields. The origin of this lies in the fact that the vector current $j_i$ can be expressed as the gradient of a symmetric tensor  current $J_{ij}$ (see (\ref{63}, \ref{64}). Thus, instead of writing the  coupling term as in  (\ref{76}), it is written as,
\begin{equation}
    {\cal L}_1= J_{ij} A_{ij} = -\delta_{ij}|\phi|^4 A_{ij}\label{a}
\end{equation}

The complete lagrangian (\ref{78}) is now expressed as,
\begin{equation}
    {\cal L} = {\cal L}_0 + {\cal L}_1 = (i\phi^{*2}\partial_i\phi\partial^i\phi - i\phi^2\partial_i\phi^*\partial^i\phi^*) - \delta_{ij}|\phi|^4 A_{ij}\label{b}
\end{equation}
Local gauge invariance{\footnote{It is actually quasi invariance since the lagrangian is invariant modulo a boundary term.}} is retained provided the tensor field transforms as,
\begin{equation}
    A_{ij}\rightarrow A_{ij} + \partial_i\partial_j \alpha\label{c}
\end{equation}
which reproduces the structure (\ref{49}).

It is also possible to establish a connection between the standard and nonstandard formulations by expressing the coupling in a dual form,
\begin{equation}
  {\cal L}_1 = j_i A_i= \partial_j J_{ij} A_i =  - \frac{1}{2} (\partial_i A_j + \partial_j A_i) J_{ij}
\end{equation}
by appropriately shifting derivatives. 

Comparison with  (\ref{a}) yields {\footnote{This relation appeared earlier in footnote 10 of \cite{S}.} },
\begin{equation}
    A_{ij}=-\frac{1}{2} (\partial_i A_j + \partial_j A_i)
    \label{81}
\end{equation}
The consistency of this identification is proved by recalling the transformation of the usual vector field (\ref{combo}) that immediately yields (\ref{c}). This gives a dual description of the theory where  standard gauge fields with minimal coupling are used or fractonic fields with nonminimal coupling{\footnote{The presence of derivatives in (\ref{81}) indicates that the coupling $J_{ij}A_{ij}$ is nonminimal.}}. Later  on we discuss the complete theory, expressed either in terms of a conventional gauge field or a fractonic field.

Let us now concentrate on the last $u$-term of the theory (\ref{58}). The current is defined in (\ref{67}, \ref{68}). Hence the first correction to the lagrangian is given by,
\begin{equation}
    {\cal L}_1= -i \partial_j\Big(\phi^2(\phi^*\partial_i\partial_j\phi^* -\partial_i\phi^*\partial_j\phi^*) -\phi^{*2}(\phi\partial_i\partial_j\phi -\partial_i\phi\partial_j\phi)\Big) A_i
    \label{82}
\end{equation}
The ensuing analysis is simplified by rephrasing the theory in terms of a fractonic field involving nonminimal coupling, following previously described steps,
\begin{equation}
    {\cal L}_1= -i\Big(\phi^2(\phi^*\partial_i\partial_j\phi^* -\partial_i\phi^*\partial_j\phi^*) -\phi^{*2}(\phi\partial_i\partial_j\phi -\partial_i\phi\partial_j\phi)\Big) A_{ij}
    \label{83}
\end{equation}
where the field $A_{ij}$ is defined in (\ref{81}). Since derivatives are involved, the iterative process continues. The new current is obtained by concentrating on the $\alpha_0$ part of (\ref{65}), so that, 
\begin{eqnarray}
    j^i  &=& \frac{\partial{\cal L}_1}{\partial(\partial_i\phi)}(i\phi) + \frac{\partial{\cal L}_1}{\partial(\partial_i\partial_j\phi)}\partial_j(i\phi) - \Big(\partial_j\frac{\partial{\cal L}_1}{\partial(\partial_i\partial_j\phi)}\Big)(i\phi)\cr &+& \frac{\partial{\cal L}_1}{\partial(\partial_i\phi^*)}(-i\phi^*) + \frac{\partial{\cal L}_1}{\partial(\partial_i\partial_j\phi^*)}\partial_j(-i\phi^*) - \Big(\partial_j\frac{\partial{\cal L}_1}{\partial(\partial_i\partial_j\phi^*)}\Big)(-i\phi^*)
    \cr
    &=& 2\partial_j(|\phi|^4 A_{ij})
    \label{84}
\end{eqnarray}
where ${\cal L}_1$ has been used from (\ref{83}). In terms of the symmetric tensor current, the result is,
\begin{equation}
    j^i= \partial_j J_{ij}\, ,\hspace{1cm} J_{ij}= 2|\phi|^4 A_{ij}
    \label{85}
\end{equation}
Thus the contribution to the lagrangian is,
\begin{equation}
   {\cal L}_2= |\phi|^4 A_{ij}^2
   \label{86}
\end{equation}
so that the functional derivative of the lagrangian (\ref{86}) reproduces the current (\ref{85}),
\begin{equation}
    J_{ij}= \frac{\delta}{\delta A_{ij}}\int{\cal L}_2
\end{equation}
\label{87}
Since no further derivatives occur, the iteration is over and the complete interacting lagrangian for the $u$- term is given by adding the contributions from (\ref{58}), (\ref{83}) and (\ref{86}),
\begin{equation}
    {\cal L}=|{\phi\partial_i\partial_j\phi- \partial_i\phi\partial_j\phi}|^2  -i\Big(\phi^2(\phi^*\partial_i\partial_j\phi^* -\partial_i\phi^*\partial_j\phi^*) -\phi^{*2}(\phi\partial_i\partial_j\phi -\partial_i\phi\partial_j\phi)\Big) A_{ij} + |\phi|^4 A_{ij}^2
    \label{88}
\end{equation}

The iterative method clearly spells out the `minimal prescription' that takes one from the original `free' theory, which here is the $u$-term in (\ref{58}), to the final form (\ref{88}). The appropriate prescription is,
\begin{equation}
\phi\partial_i\partial_j\phi- \partial_i\phi\partial_j\phi \rightarrow \phi\partial_i\partial_j\phi- \partial_i\phi\partial_j\phi -i A_{ij}\phi^2
\label{arrow}
\end{equation}
Then the pure matter sector goes to the full interacting theory (\ref{88}),
\begin{equation}
   |{\phi\partial_i\partial_j\phi- \partial_i\phi\partial_j\phi}|^2 \rightarrow |\phi\partial_i\partial_j\phi- \partial_i\phi\partial_j\phi -i A_{ij}\phi^2|^2
   \label{full}
\end{equation}
reproducing (\ref{88}). The currents in the interacting theory are likewise obtained by the same prescription (\ref{arrow}). From (\ref{68}) we obtain,
\begin{equation}
    J_{ij}\rightarrow J_{ij} + 2|\phi|^4 A_{ij} \label{681}
\end{equation}
The extra piece, expectedly, turns out to be the one found in (\ref{85}).

Gauge invariance of the lagrangian (\ref{88}) (or (\ref{arrow}) is easily seen by noting the covariant transformation property,
\begin{equation}
(\phi\partial_i\partial_j\phi- \partial_i\phi\partial_j\phi -i A_{ij}\phi^2 ) \rightarrow e^{2i\alpha(x)} (\phi\partial_i\partial_j\phi- \partial_i\phi\partial_j\phi -i A_{ij}\phi^2)
\end{equation}

The nonstandard minimal prescription (\ref{arrow}) can be related to the standard one where ordinary derivatives are replaced by covariant derivatives. To see this the double derivative term there is first symmetrised and then the minimal prescription is imposed,
\begin{eqnarray}
    \phi\partial_i\partial_j\phi&-& \partial_i\phi\partial_j\phi = \frac{1}{2}(\phi\partial_i\partial_j\phi + \phi\partial_j\partial_i\phi)- \partial_i\phi\partial_j\phi\rightarrow\cr \frac{1}{2}[\phi(\partial_i + iA_i)(\partial_j +i A_j)\phi &+& \phi (\partial_j + iA_j)(\partial_i + iA_i)\phi] - (\partial_i + iA_i)\phi(\partial_j + iA_j)\phi
\label{arrow1}
\end{eqnarray}
On simplification this expression yields,
\begin{equation}
    \phi\partial_i\partial_j\phi -\partial_i\phi \partial_j\phi \rightarrow{\phi\partial_i\partial_j\phi- \partial_i\phi\partial_j\phi} +\frac{i}{2}(\partial_i A_j + \partial_j A_i)\phi^2\label{941}
\end{equation}
which, on using (\ref{81}), is identical to (\ref{arrow}).

In terms of the vector field, therefore, the interacting lagrangian, in the standard formulation, has the structure,
\begin{equation}
    {\cal L}= |{\phi\partial_i\partial_j\phi- \partial_i\phi\partial_j\phi} +\frac{i}{2}(\partial_i A_j + \partial_j A_i)\phi^2|^2\label{94}
\end{equation}
One can also reproduce this form from the iterative Noether process by directly working with the vector gauge field.

\subsubsection{Gauge Field Lagrangian}

To obtain the complete theory one has to include the gauge field lagrangian. For the standard case, this is simply  the conventional Maxwell type term given in (\ref{20}).  Thus the complete lagrangian is given by,
\begin{equation}
   {\cal L}=  C_1 F_{0i} F_{0i} - C_2 F_{ij} F_{ij} + \frac{i}{2}\phi^*  ({\stackrel{\leftrightarrow}D_0}) \phi  + ....\label{96}
\end{equation}
and the ellipses denote terms that were calculated in the previous section  and involve both matter and gauge fields.  Specifically these do not involve the $A_0$ field{\footnote{Note that in either description, the $A_0$ field appears. It is the multiplier that enforces the relevant Gauss constraint, as shall be shown shortly.}}. Since our purpose is to calculate the gauge transformation properties, the irrelevant terms are dropped. The equation of motion for the $A_0$ field is given by,
\begin{equation}
  2C_1 \partial_i F_{0i} - \phi^*\phi = 0\label{98}
\end{equation}
The time derivative is eliminated in favour of the the canonical momenta which is defined as,
\begin{equation}
    \pi_i=\frac{\partial{\cal L}}{\partial\dot A_i}= 2 C_1 F_{0i}\label{99}
\end{equation}
so that (\ref{98}) takes the form,
\begin{equation}
    \partial_i \pi_i - \phi^*\phi = 0\label{100}
\end{equation}
The left side is just the Gauss operator that generates the standard gauge transformation (\ref{combo}),
\begin{equation}
    \delta A_i(x)= \{A_i(x) , \int_{space} \alpha\, (\partial_j \pi_j - \phi^*\phi)\} = -\partial_i\alpha(x)\label{gt}
\end{equation}

For the dual description in terms of the nonstandard 
fractonic  field $A_{ij}$, the pure gauge field lagrangian is given by,{\footnote{ Here also the coefficients of the electric and magnetic terms can be different. These are taken equal for simplifying the algebra.}}
\begin{equation}
    {\cal L}_{gauge}= \frac{1}{2}(E_{ij}^2- B_{ij}^2)\label{101}
\end{equation}
where the electric and magnetic fields are defined as,
\begin{equation}
E_{ij}= \partial_i\partial_j A_0 +\partial_0 A_{ij}\,, \hspace{1cm} B_{ij}= -\partial_i\partial_k A_{jk} + \partial_j\partial_k A_{ik}\label{102}
\end{equation}
The above lagrangian is dictated by the fact that it is the simplest generalisation of the Maxwell lagrangian that is gauge invariant under (\ref{c}) and (\ref{combo}). The complete lagrangian is therefore given by,
\begin{equation}
    {\cal L}= \frac{1}{2}(E_{ij}^2- B_{ij}^2) +\frac{i}{2}\phi^*  ({\stackrel{\leftrightarrow}D_0}) \phi
    + ....\label{103}
\end{equation}
where the ellipses denote terms containing matter and gauge fields but do not involve any $A_0$, similar to (\ref{96}). The basic difference from (\ref{96}) is in the pure gauge terms, which is expressed in terms of the vector field $A_i$. Note that while the gauge field lagrangian (\ref{101}, \ref{102}) agrees with the form given in the literature, the iterative Noether procedure is discussed in terms of the vector field and conforms to the lagrangian (\ref{96}).

The equation of motion for the $A_0$ field is given by,
\begin{equation}
    \partial_i\partial_j E_{ij} - \phi^*\phi = 0\label{104}
\end{equation}
The time derivative on $A_{ij}$ is eliminated in favour of the canonical momenta,
\begin{equation}
    \pi_{ij}= \frac{\partial{\cal L}}{\partial(\partial_0 A_{ij})} =  E_{ij}\label{105}
\end{equation}
so that (\ref{104}) is expressed as,
\begin{equation}
    \partial_i\partial_j \pi_{ij} -\phi^*\phi = 0\label{106}
\end{equation}
the left side of which is the Gauss operator. Using the usual $A_{ij} - \pi_{kl}$ Poisson algebra it is seen that this Gauss constraint generates the desired transformation (\ref{c}),
\begin{equation}
    \delta A_{ij}(x)= \{A_{ij}(x) , \, \int_{space} \alpha\,(\partial_k\partial_l \pi_{kl} -\phi^*\phi)\}= \partial_i\partial_j \alpha(x) \label{110}
\end{equation}

It is observed that the coupling of the $A_0$ and matter fields in either description remains unchanged. In both cases it acts as a lagrange multiplier that enforces the Gauss constraint. The structural similarity of the Gauss constraints (\ref{100}) and (\ref{106}) is noted. It ensures that, in both descriptions, the scalar field transforms identically as,
\begin{equation}
    \delta \phi(x)= \{\phi(x), \int_{space} \alpha\,(\partial_k\partial_l \pi_{kl} -\phi^*\phi)\} = \{\phi(x), \int_{space} \alpha\,(\partial_k \pi_{k} -\phi^*\phi)\}= i\alpha \phi(x)\label{111a}
\end{equation}
This demonstrates the validity of the choice (\ref{103}) as the lagrangan since the correct gauge transformation properties are reproduced.

\subsubsection{Extension to $SU(N)$ Invariance}

The extension to the nonabelian invariance is carried following similar steps as for conventional systems. In this case, however, the dual formulation cannot be done. Only that in terms of the standard vector field is possible. This is explicitly shown for the $t$-term in (\ref{58}). Consider therefore,
\begin{equation}
  {\cal L}_0=  (i\phi^{*2}\partial_i\phi_k\partial^i\phi_k - i\phi^2\partial_i\phi^*_k\partial^i\phi^*_k)\,\, ,\,\, \phi^2= \phi_k \phi_k
  \label{111}
\end{equation}
where $k$ runs from $1$ to the dimension of the representation of $SU(N)$ under which the scalar fields transform. The lagrangian is invariant under the global $SU(N)$ transformations (\ref{35}). The corresponding Noether currents are given by,
\begin{equation}
    j^{ia}\alpha^a = j^i= \frac{\partial{\cal L}_0}{\partial(\partial_i\phi^k)}\delta\phi^k + c.c.=2i\phi^{*2} \partial_i \phi_k (i\alpha^a(T^a)_{kj}\phi_j) + c.c \label{112}
\end{equation}
so that,
\begin{equation}
    j^{ia}= -2 (\phi^{*2} \partial_i\phi (T^{a}) \phi+ \phi^{2} \partial_i\phi^* (T^{*a}) \phi^*)\label{113}
\end{equation}
where $T^a$ denotes the $SU(N)$ generators. This current, unlike its abelian counterpart, cannot be written as the gradient of a tensor current like (\ref{67}). Hence the dual formulation in terms of a fractonic field cannot be done. Nevertheless, it is feasible to continue with the Noether process and find the gauge couplings. From (\ref{113}), the contribution to the lagrangian is written as,
\begin{equation}
    {\cal L}_1= -2 \Big(\phi^{*2} (\partial_i\phi T^a \phi) + \phi^{2} (\partial_i\phi^* T^{*a} \phi^*)\Big) A_i^a \label{114}
\end{equation}
Since there are derivative terms the iterative process continues. The next stage Noether current is obtained by replacing ${\cal L}_0$ by ${\cal L}_1$ in (\ref{112}). The result is,
\begin{equation}
    j^{ia} \alpha^a = j^i = -2 \phi^{*2} (T)^a_{kl}\phi_l i \alpha^b(T^b)_{kj} \phi_j A_i^a + c.c. \label{115}
\end{equation}
so that the current is given by,
\begin{equation}
   j^{ib}= -2i \phi^{*2} (T)^a_{kl}\phi_l (T^b)_{kj} \phi_j A_i^a + c.c. \label{116}
\end{equation}
Observing that the factor multiplying the gauge field is symmetric in $(a, b)$, the contribution to the lagrangian comes out as,
\begin{equation}
    {\cal L}_2= -i \Big(\phi^{*2} (T)^a_{kl}\phi_l (T^b)_{kj} \phi_j - \phi^{2} (T^{*a})_{kl}\phi^*_l (T^{*b})_{kj} \phi^*_j \Big)A_i^aA_i^b\label{117}
\end{equation}
The iteration terminates. The final lagrangian is given by the sum of three contributions (\ref{111}), (\ref{114}) and (\ref{117}). The result is,
\begin{equation}
     {\cal L}=  i\phi^{*2}(D_i\phi_k) (D^i\phi_k) - i\phi^2 (D_i\phi_k)^* (D^i\phi_k)^*
  \label{118}
\end{equation}
where the covariant derivatives are defined as,
\begin{equation}
    D_i\phi_k= \partial_i \phi_k + i A_i^a (T^a)_{kl}\phi_l
\end{equation}

Thus the minimally coupled form is reproduced without referring to any gauge principle. The other terms in the matter action (\ref{58}) may be similarly treated.

It is worthwhile to stress that, in the non-abelian example, only the `ordinary' $SU(N)$ transformation is a symmetry as there is no analogue of the dipolar symmetry. This is further seen from (\ref{118}) which is obtained by gauging an ordinary $SU(N)$ symmetry.

\section{Inclusion of Other Conservation Laws}

So far our analysis was confined to the conservation laws generically expressed in (\ref{52}) and (\ref{55}). Physically, these correspond to the charge (\ref{53}) and dipole (\ref{56}) symmetries, respectively. However there is a possibility that other conservation laws along these lines exist \cite{P1, P2, Umang, RB}. Thus there could be a situation where a dipole can move only normally to the dipole moment imposing further restrictions on the mobility of particles. In analogy with (\ref{56}) this is expressed by the conservation of the quantity,
\begin{equation}
    Q= \int_{space} x^2 J_0(x)
    \label{119}
\end{equation}
It is possible to verify that conservation of all three charges (\ref{53}, \ref{56}, \ref{119}) holds if the global symmetry indicated by (\ref{48}) is modified to,
\begin{equation}
    \partial_o J_0 - (\partial_i\partial_j - \frac{1}{d}\delta_{ij} \partial^2) J_{ij}=0
    \label{120}
\end{equation}
where $d$ is the spatial dimension. Now the corresponding symmetry transformation on the complex scalars takes the form (\ref{57}) where,
\begin{equation}
    \alpha(x)= \alpha_0 + \alpha_i x^i + \beta x^2\label{121}
\end{equation}
 with a constant $\beta$. A specific realisation of the matter action satisfying this property may be done but this is not necessary for the ensuing analysis.
 
 We show that once again a dual description is possible, either in terms of a standard gauge field or in terms of a symmetric traceless tensor field. A connection between these two descriptions is derived which is obviously different from the earlier example.
 
 The conservation law (\ref{120}) may be expressed in the form 
 (\ref{52}) where \cite{RB},
 \begin{equation}
   j_0=J_0 \, , \hspace{1cm}  j_i= \partial_j J_{ji} - \frac{1}{d}\partial_i J_{kk}\label{122}
 \end{equation}
 The temporal component remains identical to the earlier model. So we concentrate on the spatial current. Coupling with the standard $A_i$ field may be rephrased in favour of a symmetric  tensor field as,
 \begin{equation}
     j_i A_i = J_{ij} A_{ij} \label{123}
 \end{equation}
 where,
 \begin{equation}
     A_{ij}= -\frac{1}{2}(\partial_i A_j + \partial_j A_i -\frac{2}{d}\delta_{ij} \partial_l A_l)\label{124}
 \end{equation}
 
 The condition (\ref{123}) is general and is a manifestation of the dual nature of the fields. Any change in the relation between the vector and tensor currents is precisely compensated by a corresponding change in the relation between the vector and tensor gauge fields, such that the condition (\ref{123}) is preserved. For instance, in the present case, (\ref{122}) and (\ref{124}), respectively, give the relations between the currents and the gauge fields. Similarly, for the earlier case involving the charge and dipole symmetry only, the relation among the currents is given by (\ref{51}) while that between the gauge fields is (\ref{81}). It is easy to verify that, given any one relation among (\ref{122}) and (\ref{124}), the other can be derived by exploiting (\ref{123}). As an example, contracting (\ref{124}) by $J_{ij}$, appropriately shifting the derivatives, using (\ref{123}), leads to (\ref{122}).

 The tensor field in (\ref{124}) is traceless $(A_{ii}=0)$ and is a consequence of the modified symmetry.
 Using (\ref{combo}) the transformation of the above tensor field is given by,
 \begin{equation}
     \delta A_{ij}= \partial_i\partial_j\alpha - \frac{1}{d} \delta_{ij}\partial^2\alpha\label{126}
 \end{equation}
 
 While the magnetic field (\ref{102}) remains gauge invariant under this new transformation, the electric field is not. To make it gauge invariant, the definition of the electric is modified to,
 \begin{equation}
E_{ij}= (\partial_i\partial_j - \frac{1}{d}\delta_{ij}\partial^2)A_0 +  \partial_0 A_{ij}\,, \hspace{1cm} B_{ij}= -\partial_i\partial_k A_{jk} + \partial_j\partial_k A_{ik}\label{127}
\end{equation}
 The lagrangian of the theory is defined by (\ref{103}) with the electric and magnetic fields given above. The pure matter sector naturally is different but its explicit form is not required here. We want to compute the Gauss constraint from this lagrangian and deduce the transformation  law (\ref{126}). The equation of motion for the $A_0$ field is altered from (\ref{104}) to,
 \begin{equation}
    (\partial_i\partial_j -\frac{1}{d} \delta_{ij} \partial^2) E_{ij} - \phi^*\phi = 0\label{128}
\end{equation}
 Since the momenta has the same functional form as (\ref{105}), the Gauss constraint follows from (\ref{128}),
 \begin{equation}
    (\partial_i\partial_j -\frac{1}{d} \delta_{ij} \partial^2) \pi_{ij} - \phi^*\phi = 0\label{129}
\end{equation}
 which reproduces the transformation (\ref{126}),
 \begin{equation}
     \delta A_{ij}= \{A_{ij}(x) , \, \int_{space} \alpha [\,(\partial_k\partial_l - \frac{1}{d}\delta_{kl}\partial^2)\pi_{kl} -\phi^*\phi]\}=  (\partial_i\partial_j - \frac{1}{d} \delta_{ij}\partial^2)\alpha\label{130}
 \end{equation}
 while the transformation of the scalar field remains same as before. This shows that the suggested lagrangian yields an appropriate generator (Gauss constraint) that generates the desired transformation laws. The formulation in terms of the standard gauge field is carried out as usual with a Maxwell type term (see (\ref{96})).
 
 \section{Conclusions}

 The gauge principle, which requires the invariance of a physical theory under local gauge transformations, is a cornerstone in the development of modern field theory and gravity, both relativistic and nonrelativistic \cite{hooft}. It has also formed the basis for introducing interactions in a free theory of matter invariant under global symmetry transformations. Ordinary derivatives are replaced by covariant derivatives such that the additional connection term compensates for the difference in the transformation of the field at two neighbouring points. Field strengths and/or curvatures are now defined using the Ricci identity and the action can be written. Thus the origins of the gauge principle are more geometrical than physical.
 
 An alternative and more physical approach was presented by Deser \cite{Deser} where interactions, instead of the gauge principle, play the pivotal role. It is based on an iterative Noether prescription where  interactions are introduced in a step by step process. The term found from the original source is used to calculate the source for the next step and so on till the iteration terminates when no new source is obtained. This method is therefore a bridge between the two theorems of Noether. Starting from a theory that has a global symmetry (Noether's first theorem), the iteration eventually yields an interacting theory invariant under local symmetry transformations (Noether's second theorem).

  
  So far the iterative Noether scheme was applied to usual (non higher derivative) systems \cite{DHB, AS, DH, K, K1, BMM}. Here we presented an application to higher derivative theories invariant under both charge and dipole symmetries. 
   Such theories have been discussed recently, both on general terms \cite{S} as well as by concrete examples \cite{S, Pretko, RB}. They have generated considerable interest since they predict a new type of topological matter, called fractons, that have restricted mobility. The iterative prescription, which begins from the free (quadratic) theory, has to be suitably generalised to account for the fact that the new  `free' matter theory is no longer quadratic. Also,  in the present case, the presence of   higher derivatives is an additional complication. The application of the present scheme yields a dual picture which is traced to the presence of multiple global symmetries. 
   
   The iterative Noether prescription allows the introduction of interactions in two distinct ways, the origin of which is in the dual picture stated above.  In one case they are inducted  by a  vector field and the additional terms can be written so that the complete theory is the standard minimally coupled theory, where ordinary derivatives are replaced by their covariant expressions. Alternatively, interactions may also be expressed 
    in terms of a symmetric tensor field recently encountered in fractonic theories. In fact this method of introducing interactions naturally led to a nonstandard `minimal prescription' where the  complete theory is expressed by replacing the ordinary double derivatives by appropriate `covariant' expressions involving the tensor fields (see (\ref{arrow}). Furthermore, an equivalence of these two minimal coupling prescriptions  is shown by exploiting a map that relates the vector and tensor gauge fields. 
    
    The analysis also dictates the structure of the pure gauge term in the lagrangian. Self consistency is established from the constraint structure of the total lagrangian including gauge, matter and interacting sectors. Extension to other fractonic theories with more conservation laws following from additional global symmetries is shown. 
  
  It would be desirable to apply the methods formulated here to an analysis of the current algebra in these theories. As we mentioned, the presence of multiple global symmetries yields a dual description leading to distinct currents. Their algebra, with or without the inclusion of gauge fields, could possibly provide new sights into the structure of these theories. 
  \smallskip
  
  Acknowledgements: This work was supported by a (DAE) Raja Ramanna Fellowship.


\begin{thebibliography}{999}

\bibitem{P1} M. Pretko, Subdimensional particle structure of higher rank U(1) spin liquids, Phys. Rev. B
95, 115139 (2017), arXiv:1604.05329.

\bibitem{P2} M. Pretko, Generalized electromagnetism of subdimensional particles: A spin liquid story,
Phys. Rev. B 96, 035119 (2017), arXiv:1606.08857

\bibitem{DRH} O. Dubinkin, A. Rasmussen, and T. L. Hughes, Higher-form gauge symmetries in multipole
topological phases, Ann. Phys. (N.Y.) 422, 168297 (2020), arXiv:2007.05539.

\bibitem{S} N. Seiberg, Field theories with a vector global symmetry, SciPost Phys. 8, 050 (2020),
arXiv:1909.10544.

\bibitem{C} C. Chamon, Quantum glassiness in strongly correlated clean systems: An example of topological overprotection, Phys. Rev. Lett. 94, 040402 (2005), cond-mat/0404182.

\bibitem{NH} 
R. M. Nandkishore and M. Hermele, Fractons, Ann. Rev. Condensed Matter Phys. 10
(2019) 295–313, [arXiv:1803.11196], and references therein.

\bibitem{PCY}
M. Pretko, X. Chen, and Y. You, Fracton Phases of Matter, Int. J. Mod. Phys. A 35
(2020), no. 06 2030003, [arXiv:2001.01722].

\bibitem{QRH} M. Qi, L. Radzihovsky, and M. Hermele, Fracton phases via exotic higher-form symmetry breaking, Ann. Phys. (N.Y.) 424, 168360 (2021), arXiv:2010.02254.

\bibitem{SS}  N. Seiberg and S.-H. Shao, Exotic symmetries, duality, and fractons in 2+1-dimensional quantum field theory, SciPost Phys. 10, 027 (2021), arXiv:2003.10466.

 

\bibitem{SPP} Kevin Slagle, Abhinav Prem, Michael Pretko
Symmetric Tensor Gauge Theories on Curved Spaces, Annals of Physics 410 (2019) 167910, arXiv:1807.00827.

\bibitem{Umang} 
Yi-Hsien Du, Umang Mehta, Dung Xuan Nguyen, and Dam Thanh Son
Volume-preserving diffeomorphism as nonabelian higher-rank
gauge symmetry, arXiv:2103.09826.

\bibitem{RB} R. Banerjee, Hamiltonian formulation of higher rank symmetric fractonic gauge theories, arXiv:2105.04152; Eur. Phys. J. C (2022) 82:22.

\bibitem{Pretko} M. Pretko, The fracton gauge principle, Phys. Rev. B98, 115134 (2018).

\bibitem{Deser} S. Deser, Self-interaction and gauge invariance, Gen. Rel. Grav. 1, 9, (1970); Since this paper is not readily available, Deser subsequently posted an arXiv: gr-qc/0411023 (V3) which is basically the original version.

\bibitem{DHB} S. Deser, J.H. Kay and D.G. Boulware, Supergravity from selfinteraction, Physica (Amsterdam) 96A, 141 (1979).

\bibitem{AS} C. Aragone and J. Stephany, Nonabelian Chern-Simons topological coupling from selfinteraction, Rev. Bras. Fis. 16, 287, (1986).

\bibitem{DH} S. Deser and M. Henneaux, Gauge properties of conserved currents in abelian versus nonabelian theories, Mod. Phys. Letts. A10, 991 (1995).

\bibitem{K} A. Khoudeir, Nonabelian antisymmetric vector coupling from selfinteraction, Mod. Phys. Letts. A11, 2489 (1996).

\bibitem{K1} A. Khoudeir, Nonabelian self duality from selfinteraction, Mod. Phys. Letts. A16, 2123 (2001).

\bibitem{BMM} A. Basu, P. Majumdar and I. Mitra, Gauge invariant matter field actions from an iterative Noether coupling,  Phys. Rev. D98, 105018 (2018).

\bibitem{hooft} G. 't Hooft, {\it{Under the Spell of the Gauge Principle}}, World Scientific Publishing Company, (1994).

\end{thebibliography}
\end{document}